\documentclass[twocolumn,amsmath,amssymb,nofootinbib,superscriptaddress,floatfix]{revtex4-2}
\usepackage[utf8]{inputenc}
\usepackage[pdftex]{graphicx}
\usepackage{mathrsfs}
\usepackage[colorlinks, breaklinks, urlcolor={blue}, linkcolor={blue}, citecolor={blue}]{hyperref}
\usepackage{array}
\usepackage{amsmath}
\usepackage{type1cm}
\usepackage[english]{babel}
\usepackage{lmodern}
\usepackage{microtype}
\usepackage{booktabs}
\usepackage{braket}
\usepackage{xcolor}
\usepackage{orcidlink}
\usepackage{tikz}
\usepackage{quantikz}

\begin{document}
\title{Maximally Sliced States as a Framework for Quantum Coherence, Entanglement, and Teleportation under Decoherence}

\author{Anushree Pandey \orcidlink{0000-0002-0655-1773}}
\email[]{a.bhattacharya.tmsl@ticollege.org}
\affiliation{Department of Mathematics, Techno Main Salt Lake (Engg. Colg.), \\Techno India Group, EM 4/1, Sector V, Salt Lake, Kolkata  700091, India}

\author{Sovik Roy\orcidlink{0000-0003-4334-341X}}
\email[]{s.roy2.tmsl@ticollege.org}
\affiliation{Department of Mathematics, Techno Main Salt Lake (Engg. Colg.), \\Techno India Group, EM 4/1, Sector V, Salt Lake, Kolkata  700091, India}

\begin{abstract}
\noindent Quantum resources such as coherence and entanglement need not exhibit the same dependence on the parameters of a quantum state or respond similarly to environmental noise. We investigate these relationships using the three-qubit maximally sliced (MS) state, whose continuous parameter provides a tunable framework for redistributing quantum correlations. We evaluate the relative entropy of coherence, $\ell_1$-norm coherence, and basis-independent coherence and compare them with the three-tangle. After tracing out one qubit, we analyze the coherence, concurrence, and optimal teleportation fidelity of the resulting bipartite state. The concurrence is found to coincide with the $\ell_1$-norm coherence because both are determined by the same off-diagonal matrix elements. We then study the effects of amplitude damping and dephasing noise on the bipartite maximally sliced state derived from the tripartite form of maximally sliced state. Although the two channels modify the MS state through different physical mechanisms, the teleportation fidelity $f_{\mathrm{opt}}^T$ in the regime of nonzero concurrence satisfies the condition $f_{\mathrm{opt}}^T=\frac{2+\mathcal{C}}{3}$ for both channels. The distinct dependence of concurrence and coherence on the two noise parameters demonstrates the usefulness of the MS family as a unified framework for studying quantum resources and teleportation under decoherence.
\end{abstract}


\maketitle

\section{Introduction} 

\noindent Quantum entanglement is one of the most distinctive resources of quantum mechanics and plays a central role in quantum information processing \cite{horo2009,roy2020}, including quantum communication \cite{gisin2007,cg2016}, quantum computation \cite{dp1995,nc2000}, quantum cryptography \cite{bb84,bb92,pira2020}, and quantum teleportation \cite{bb1993,bouw1997,roy2016}. While bipartite entanglement has been extensively investigated \cite{sv2009,gg2021,dd2022}, multipartite entanglement possesses a richer structure because the correlations may be distributed among several subsystems in inequivalent ways \cite{thap1999,coff2000,m2016}. In particular, the relationship between genuine multipartite entanglement, pairwise entanglement, quantum coherence, and the performance of quantum-information protocols remains an important subject of investigation. A major difficulty in such studies is that different quantum resources need not vary in the same manner, and consequently a change in one resource does not necessarily imply a corresponding change in another.\\

\noindent Among three-qubit entangled states, the maximally sliced (MS) state occupies a particularly interesting position. The MS state was introduced in the context of the classification of three-qubit pure states and provides a parametrized family of states belonging to the $GHZ$ class \cite{sud2000,roy2018}. Its importance arises from the fact that a single continuous parameter allows the distribution of quantum correlations among the three constituent qubits to be varied in a controlled manner. Thus, rather than comparing a few isolated representative states, such as the $GHZ$ and $W$ states, the MS family provides a continuous framework for examining how different manifestations of quantum correlations evolve as the state is continuously deformed. This tunability makes the MS state particularly suitable for studying the interplay between genuine tripartite entanglement and correlations retained between selected pairs of qubits.\\

\noindent The tripartite maximally sliced state is defined as 
\begin{eqnarray}
\label{msstate}
|\Psi_{MS}\rangle = \frac{1}{\sqrt{2}}\Big\lbrace |000\rangle + \cos(\theta)|110\rangle + \sin(\theta)|111\rangle\Big\rbrace,
\end{eqnarray}
where $\theta$ is the parameter of the state.\\

\noindent The MS state is also especially useful because its limiting cases provide physically distinct correlation structures within the same family. For the parametrization considered here, the state continuously connects a configuration in which one of the qubits is separable from an entangled pair to the $GHZ$ type limit. Consequently, tracing out one of the parties does not merely produce an arbitrary mixed two-qubit state; it gives a well-defined bipartite channel whose properties can be followed continuously as the parameter of the original tripartite state is varied. This feature provides a natural bridge between multipartite entanglement and bipartite quantum-information tasks. In particular, it allows one to ask how genuine tripartite correlations in the parent state are related to the entanglement and operational usefulness of the reduced two-qubit state. The MS family provides a convenient analytical setting in which multipartite entanglement, coherence, and operational performance can be examined within a common parametrized framework. Nevertheless, a systematic comparison of different coherence measures, their relation to genuine tripartite entanglement, and the way these quantities are connected with the entanglement and teleportation capability of the reduced bipartite state requires further attention.\\

\noindent Quantum coherence is particularly relevant in this context. Coherence is the manifestation of quantum superposition in a chosen reference basis and constitutes a fundamental quantum resource in its own right \cite{as2017,tb2014,roy2023}. Conventional measures such as the relative entropy of coherence and the $\ell_1$-norm coherence are basis dependent and therefore describe how coherence is manifested relative to a specified basis. On the other hand, basis-independent measures of coherence provide a complementary description of the intrinsic coherent content of a quantum state. The simultaneous consideration of these measures is especially informative for multipartite states, where a variation in basis-dependent coherence need not signify a corresponding variation in the underlying quantum resource. The MS family, with its continuously tunable parameter, offers a particularly transparent setting for examining this distinction.\\

\noindent An additional motivation comes from the relation between coherence and entanglement. Although coherence is a more fundamental resource that can exist even in a single quantum system, entanglement represents a distinctly nonlocal form of quantum correlation between subsystems \cite{horo2009}. There is therefore no general reason for different coherence measures and entanglement measures to exhibit identical parameter dependence. The MS state provides a useful analytical example in which this relationship can be investigated explicitly. On the other hand, the reduced bipartite MS state also provides a natural connection to quantum teleportation. After one party is traced out, the remaining two-qubit state can serve as a quantum channel, and its usefulness can be quantified by calculating the optimal teleportation fidelity of the two-qubit state. This is another reason as to why MS family to be examined from both a structural and an operational perspective.\\

\noindent In realistic quantum-information settings, however, quantum resources are inevitably affected by interactions with the environment. Decoherence can modify both the coherent structure and the entanglement of a quantum state, thereby reducing its usefulness for information-processing tasks. Different noise mechanisms affect a quantum state in fundamentally different ways. Amplitude damping represents energy relaxation and changes both the populations and the off-diagonal elements of the density matrix, whereas dephasing primarily suppresses relative phase coherence while leaving the populations unchanged. Comparing these two channels for the same bipartite state derived from tripartite MS state therefore provides a direct way of distinguishing the effects of population relaxation from those associated with phase decoherence.\\

\noindent In the present work we undertake a systematic study of the MS state. We first investigate the relative entropy of coherence, $\ell_1$-norm coherence, and basis-independent coherence of the tripartite MS state and examine their dependence on the state parameter. We then compare these coherence measures with the three-tangle in order to distinguish basis-dependent manifestations of coherence from genuine tripartite entanglement. Next, by tracing out one party, we investigate the resulting bipartite MS state and establish the relation between its coherence, concurrence, and teleportation fidelity. Finally, we examine the evolution of the bipartite state under the noises such as amplitude damping and dephasing and compare the responses of the different coherence measures, concurrence, and teleportation fidelity, when the bipartite state is subjected to the two distinct noise mechanisms.\\

\noindent The organization of the manuscript is as follows. Sec.~\ref{subsec:generation_MS} presents a quantum-circuit realization of the maximally sliced (MS) state and provides an operational interpretation of the state parameter \(\theta\). Sec.~\ref{sec:basics} introduces the quantum-resource measures employed in this work, including the relative entropy of coherence, \(\ell_1\)-norm coherence, basis-independent coherence, concurrence, and teleportation fidelity. Sec.~\ref{sec:qrms} presents the results for the MS state in the absence of noise, beginning with the tripartite state and subsequently considering the reduced bipartite state obtained by tracing out one party. Sec.\ref{sec:adc} investigates the effects of amplitude damping on the bipartite MS state, with particular emphasis on the evolution of coherence, concurrence, and teleportation fidelity. Sec.\ref{sec:dp} examines the corresponding effects of dephasing on the same bipartite state derived from tripartite MS state and consequently compare the resulting behavior with that obtained under amplitude damping in sec.~\ref{sec:adc}. Finally, the main findings and their physical implications are summarized in the concluding sec.~\ref{sec:conc}.\\

\section{Quantum Circuit Realization of Maximally Sliced State}
\label{subsec:generation_MS}

\noindent The one-parameter structure of the maximally sliced (MS) state also allows for a simple quantum-circuit realization. Starting from the three-qubit computational basis state $|000\rangle$, the MS state can be generated using a Hadamard gate, a controlled-NOT (CNOT) gate, and a controlled single-qubit rotation. In particular, the preparation circuit is shown in Figure.~\ref{fig:MS_generation}. Initially, a Hadamard gate is applied to the first qubit, giving
\begin{equation}
|000\rangle
\xrightarrow{H_1}
\frac{1}{\sqrt{2}}
\left(|000\rangle+|100\rangle\right).
\end{equation}
\noindent A CNOT gate with the first qubit as the control and the second qubit as the target then produces
\begin{equation}
\frac{1}{\sqrt{2}}
\left(|000\rangle+|110\rangle\right).
\end{equation}
\noindent Finally, a controlled rotation $R_y(2\theta)$ is applied to the third qubit, with the first qubit serving as the control. The action of the rotation on the target qubit is
\begin{equation}
R_y(2\theta)|0\rangle
=
\cos\theta\,|0\rangle+
\sin\theta\,|1\rangle.
\end{equation}
\noindent Since the rotation is activated only when the first qubit is in the state $|1\rangle$, the resulting state is
\begin{equation}
\begin{aligned}
|\Psi_{\rm MS}\rangle
&=
\frac{1}{\sqrt{2}}
\left[
|000\rangle+
|11\rangle
\left(
\cos\theta\,|0\rangle+
\sin\theta\,|1\rangle
\right)
\right] \\
&=
\frac{1}{\sqrt{2}}
\left(
|000\rangle+
\cos\theta\,|110\rangle+
\sin\theta\,|111\rangle
\right).
\end{aligned}
\label{eq:MS_preparation}
\end{equation}

\noindent Thus, the parameter $\theta$ has a direct operational interpretation as half of the rotation angle of the controlled $R_y$ gate. Varying $\theta$ continuously therefore generates the entire family of MS states. In particular, at $\theta=0$ the MS state reduces to $\frac{|000\rangle+|110\rangle}{\sqrt{2}}$, corresponding to a Bell state of the first two qubits with the third qubit separable, whereas at $\theta=\frac{\pi}{2}$ it reduces to the $GHZ$ state $\frac{|000\rangle+|111\rangle}{\sqrt{2}}$. This preparation circuit highlights the tunable character of the MS family and provides an operational basis for studying its coherence, entanglement, and teleportation properties under different noise channels.

\begin{figure}[htbp]
    \centering
    \begin{quantikz}[row sep={0.55cm,between origins},
                     column sep=0.55cm]
        \lstick{$q_1$} & \gate{H} & \ctrl{1} & \ctrl{2} & \qw \\
        \lstick{$q_2$} & \qw      & \targ{}  & \qw      & \qw \\
        \lstick{$q_3$} & \qw      & \qw      & \gate{R_y(2\theta)} & \qw
    \end{quantikz}
    \caption{Quantum circuit for generating the three-qubit maximally
    sliced state.}
    \label{fig:MS_generation}
\end{figure}

\section{Quantification of Quantum Resources and Teleportation Performance}
\label{sec:basics}
\noindent In this section we discuss the measures such as relative entropy of coherence, $\ell_1$-norm of coherence, basis independent coherence, concurrence and teleportation fidelity. In the present work, these measures are applied to the state under consideration before the state is subjected to the noise as well as to the post-noise scenario of the state. We first discuss these measures.\\

\noindent \textit{Quantum coherence:} The manifestation of the superposition principle is captured by quantum coherence, which is characterized by the presence of the off-diagonal elements when the given quantum state is expressed in its density-matrix formalism. Among various measures of quantum coherence, \textit{relative entropy of coherence} \cite{tb2014} is often used. For a state $\rho$, the relative entropy of coherence is defined as 
\begin{eqnarray}
\label{cr1}
\mathcal{C}_{re}(\rho) = \min_{\sigma\in \mathcal{I}}S(\rho\|\sigma),
\end{eqnarray}
\noindent $S$ being the von-Neumann entropy. Here the minimum is taken over the set of the incoherent states $\mathcal{I}$ which are the states without quantum coherence. The computational formula for calculating $\mathcal{C}_{re}(\rho)$ also can be derived from Eq.(\ref{cr1}) and is given by \cite{tb2014}
\begin{eqnarray}
\label{cr2}
\mathcal{C}_{re}(\rho) = S(\rho_d) - S(\rho).
\end{eqnarray}
\noindent Here $\rho_d$ is the dephased state in the reference basis $\lbrace |~i~\rangle\rbrace$, i.e. the state obtained from $\rho$ by deleting all the off-diagonal entries. The von-Neuman entropy $S$ of state $\rho$ is
\begin{eqnarray}
S(\rho) = -\sum_{x}\lambda_x~\log_2(\lambda_x),
\end{eqnarray}
\noindent $\lambda_x$ are the eigenvalues of the state $\rho$.\\

\noindent Another measure to quantify the coherence, known as $\ell_1$-norm of coherence \cite{tb2014}.  Since the 
quantum coherence is a basis dependent quantity, we fix the reference basis  $\ket{i}$ for a given quantum state.    The $\ell_{1}-$ norm of coherence is  defined 
as
\begin{eqnarray}
\label{l1norm}
\mathcal{C}_{\ell_{1}}(\rho) = \sum_{i,j; i\ne j} \vert \rho_{ij}\vert,
\end{eqnarray}
\noindent where $\rho_{ij} = \langle i\vert \rho\vert j\rangle$ is the matrix element corresponding to the $i^{th}$ row and $j^{th}$ column.\\

\noindent Although both $\mathcal{C}_{re}$ and $\mathcal{C}_{\ell_{1}}$ quantify quantum coherence with respect to the same reference basis, they are inequivalent coherence measures and, in general, need not yield identical numerical values. The measure $\mathcal{C}_{\ell_{1}}$ quantifies coherence through the sum of the absolute values of the off-diagonal elements of the density matrix, whereas $\mathcal{C}_{re}$ quantifies coherence in terms of the relative entropy between the quantum state and its diagonal part. Consequently, even for the same quantum state and reference basis, the two measures can exhibit different numerical values and functional dependence on the state parameters.\\

\noindent Radhakrishnan \textit{et al.} has proposed the measure of basis-independent coherence \cite{radha2019}. This measure is different from the relative entropy of coherence in the sense that in calculating relative entropy of coherence one needs to minimize the distance of the given state $\rho$ from the set of all states $\sigma$ present in the incoherent set $\mathcal{I}$ as shown in Eq.(\ref{cr1}), while the calculation of basis independent coherence only needs to take the maximally mixed state as the reference state and hence the minimization need not be done. Moreover, the set of incoherent states is dependent upon a particular choice of basis and so relative entropy of coherence is a basis dependent measure whereas the basis-independent measure does not require this constraint. The basis-independent measure by Radhakrishnan \textit{et al.} \cite{radha2019}, which we name here as \textit{BICoher} and denote by $\mathcal{C}_{BI}$, is defined as
\begin{eqnarray}
\label{BI}
\mathcal{C}_{BI}(\rho) = \sqrt{S\Big(\frac{\rho + \frac{I}{d}}{2}\Big) - \frac{S(\rho) + \log_{2}d}{2}},
\end{eqnarray}
\noindent where  $d$ is the dimension of the system, $\frac{I}{d}$ is the maximally mixed state.\\

\noindent \textit{Concurrence:} Non-local correlations between quantum systems give rise to the important quantum resource called entanglement. Unlike coherence, which can exist at the single-qubit level, entanglement requires more than one qubit in the quantum system. Entanglement as a resource is quantifiable. Measures like relative entropy of entanglement, negativity etc are often used to quantify it, however, \textit{concurrence} is the most prominent measure for quantifying entanglement of two-qubit systems. It is to be noted that concurrence works equally well for both the pure as well as the mixed states. Proposed by Wootters, the concurrence of a quantum state $\rho$ is given by \cite{wootters1998,wootters2001}
\begin{eqnarray}
\label{co1}
\mathcal{C}(\rho) = \max \lbrace ~ 0,~\sqrt{\lambda_1}-\sqrt{\lambda_2}-\sqrt{\lambda_3}-\sqrt{\lambda_4}~\rbrace,
\end{eqnarray}
\noindent where $\lambda_1\geq \lambda_2 \geq \lambda_3 \geq \lambda_4$ are the eigenvalues of the matrix $\rho\tilde{\rho}$. The spin-flipped density matrix $\tilde{\rho}$ is 
\begin{eqnarray}
\label{cospin}
\tilde{\rho} = (\sigma_y \otimes \sigma_y)\rho^*(\sigma_y \otimes \sigma_y),
\end{eqnarray}
\noindent where $\sigma_y$ is the Pauli matrix in the $y$ direction and $\tilde{\rho}$ is in the same basis as $\rho$,  $\rho^*$ is the complex conjugate of the density matrix $\rho$.\\

\noindent There is another fundamental technical difference between coherence and concurrence. While concurrence ($\mathcal{C}$) is a basis-independent measure, most of the measures of coherence and obviously the relative entropy of coherence ($\mathcal{C}_{re}$), is a basis-dependent measure \cite{tb2014}. This means if one measures $\mathcal{C}_{re}$ in either computational basis or in say, Bell-basis, in both the cases the value of the coherences would be different). However, \textit{BICoher} of coherence, unlike other measures of coherence, defined in Eq.~(\ref{BI}) is basis-independent measure.\\

\noindent Now we discuss the measure which validates a given quantum state with respect to its capability of teleporting unknown quantum state and is captured by the quantity known as teleportation fidelity.\\

\noindent \textit{Teleportation fidelity:}
\noindent A quantum channel is useful for teleportation only if its fidelity exceeds $\frac{2}{3}$ which is the maximum possible fidelity achievable by means of Local Operations and Classical Communication (LOCC) and is known as the classical fidelity \cite{pop1994,pop1995,gisin1996}. For an arbitrary density matrix $\rho$, the optimal teleportation 
fidelity is 
\begin{eqnarray}
\label{telepfid1}
f^{T}_{opt}(\rho) = \frac{1}{2}\Bigg[1 + \frac{N(\rho)}{3}\Bigg],
\end{eqnarray} 
where $N(\rho) = \sum_{i=1}^{3}\sqrt{u_{i}}$. Here, $u_{i}$'s are the eigenvalues of the matrix $T^{\dagger}T$. The elements of $T$,  denoted by $t_{ij}$ are
\begin{eqnarray}
\label{telepfid2}
t_{ij} = Tr(\rho\sigma_{i}\otimes \sigma_{j}),
\end{eqnarray}
and $\sigma_{k}$'s denote the Pauli spin matrices. In terms of the quantity $N(\rho)$, one can say that any mixed state $\rho$ is useful for standard teleportation if and only if $N(\rho) > 1$ \cite{horotele1996,horoinfo1996}  .\\

\noindent We next proceed for \textit{pre and post noise analysis} of the tripartite maximally sliced state under consideration.

\section{Quantum Resources of the Maximally Sliced State}
\label{sec:qrms}
\noindent The density matrix representation corresponding to the state defined in Eq.(\ref{msstate}) is given by
\begin{equation}
\label{densityMS}
\rho^{MS} =
\left(
\begin{array}{cccccccc}
\frac{1}{2} & 0 & 0 & 0 & 0 & \frac{\cos\theta}{2} & 0 & \frac{\sin\theta}{2} \\
0 & 0 & 0 & 0 & 0 & 0 & 0 & 0 \\
0 & 0 & 0 & 0 & 0 & 0 & 0 & 0 \\
0 & 0 & 0 & 0 & 0 & 0 & 0 & 0 \\
0 & 0 & 0 & 0 & 0 & 0 & 0 & 0 \\
\frac{\cos\theta}{2} & 0 & 0 & 0 & 0 & \frac{\cos^2\theta}{2} & 0 & \frac{\cos\theta\sin\theta}{2} \\
0 & 0 & 0 & 0 & 0 & 0 & 0 & 0 \\
\frac{\sin\theta}{2} & 0 & 0 & 0 & 0 & \frac{\cos\theta\sin\theta}{2} & 0 & \frac{\sin^2\theta}{2}
\end{array}
\right).
\end{equation}
\noindent Coffman \textit{et al.} proposed the measure of genuine tripartite entanglement, called the \textit{three-tangle} \cite{coff2000} and this is given by,
\begin{eqnarray}
\label{threetangle}
\tau = \mathcal{C}^2_{A(BC)} - \mathcal{C}^2_{AB} - \mathcal{C}^2_{AC},
\end{eqnarray}
\noindent where $\mathcal{C}^2$ is the concurrence-squared. Here $\mathcal{C}^2_{A(BC)}$ measures entanglement between qubits $A$ (Alice) and the joint state of qubits $B$ (Bob) and $C$ (Charlie) while, $\mathcal{C}^2_{AB}$ (and $\mathcal{C}^2_{AC}$) measures the entanglement between qubits $A$ and $B$ (and $C$). Using Eq.(\ref{threetangle}), the three-tangle of state $\rho^{MS}$ is found to be
\begin{eqnarray}
\label{tanglerhoMS}
\tau(\rho^{MS}) = \sin^2(\theta).
\end{eqnarray}
\subsection{Pre-Noise Scenario:}
\subsubsection{Tripartite MS state:}
\noindent Using Eqs.~(\ref{cr2}), (\ref{l1norm}) and (\ref{BI}), the relative entropy of coherence, $\ell_{1}$-norm, and basis independent coherence of the state $\rho^{MS}$ of Eq.(\ref{msstate}) is given by

\begin{eqnarray}
\label{allcoherences}
\mathcal{C}_{re}(\rho^{MS}) &=& 1 - \sin^{2}(\theta)~\log_{2}\sin(\theta) - \cos^{2}{\theta}\log_{2}\cos(\theta),\nonumber\\
\mathcal{C}_{\ell_{1}}(\rho^{MS}) &=& \cos(\theta) + \sin(\theta) + \frac{\sin(2\theta)}{2},\nonumber\\
\mathcal{C}_{BI}(\rho^{MS}) &\approx & 0.84671, ~~~\text{independent of $\theta$}.
\end{eqnarray}

\noindent The relative entropy of coherence and $\ell_{1}$-norm have been calculated with respect to the computational basis. We now plot these expressions in the following figure.\\

\begin{figure}[htbp]
    \centering
    \includegraphics[width=0.98\columnwidth]{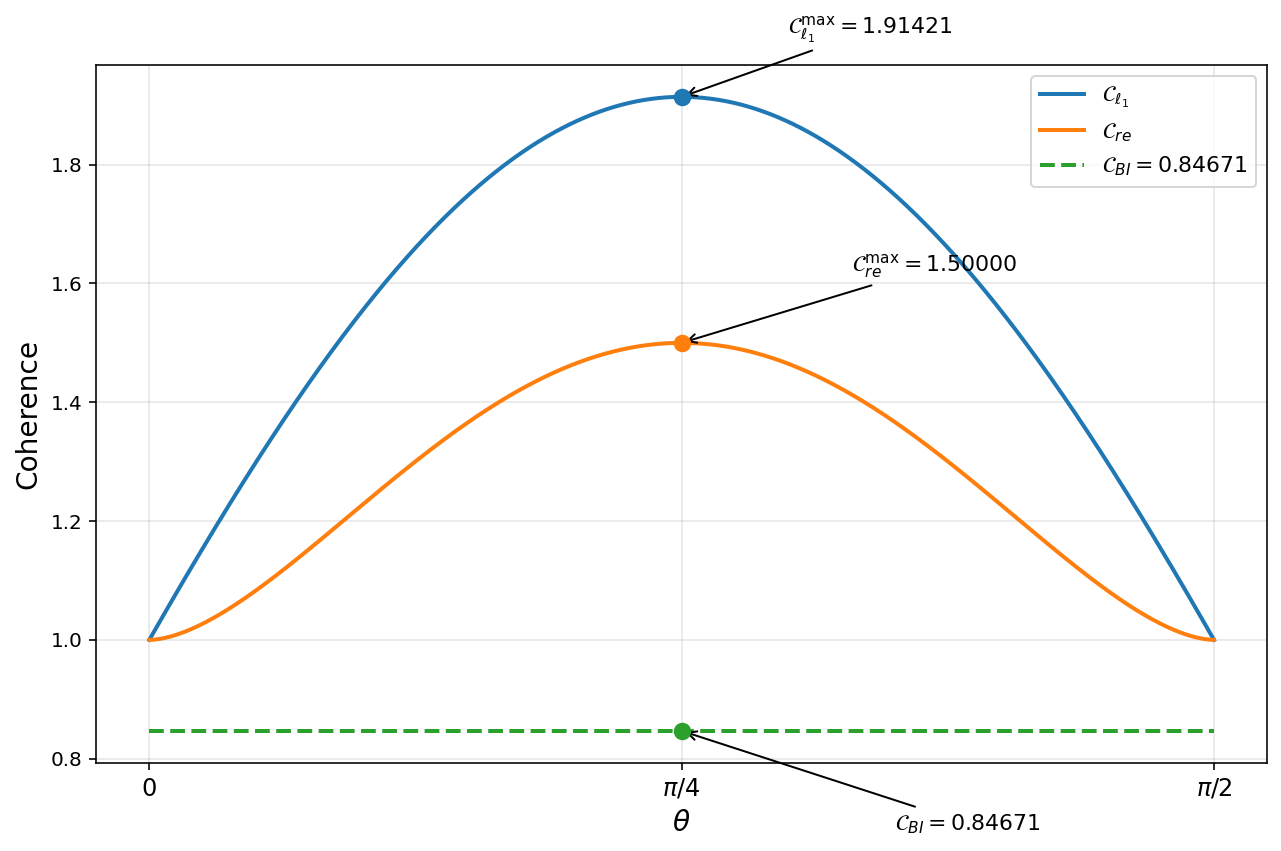}
    \caption{Relative entropy, $\ell_1$-norm, and basis-independent coherence of the tripartite MS state as functions of $\theta$.}
    \label{fig1}
\end{figure}

\noindent Figure~\ref{fig1} compares the relative entropy of coherence $\mathcal{C}_{re}$, the $\ell_1$-norm of coherence $\mathcal{C}_{\ell_1}$, and the basis-independent coherence $\mathcal{C}_{BI}$ for the tripartite maximally sliced (MS) state of Eq.(\ref{msstate}) as a function of $\theta$. Both $\mathcal{C}_{re}(\rho^{MS})$ and $\mathcal{C}_{\ell_1}(\rho^{MS})$ exhibit a pronounced non-monotonic dependence on $\theta$ and are symmetric about $\theta=\frac{\pi}{4}$. Specifically, these two measures attain their respective maxima at the balanced point $\theta=\frac{\pi}{4}$, where $\mathcal{C}_{\ell_{1}}(\rho^{MS}) =\sqrt{2}+\frac{1}{2} \approx 1.9142$ and $\mathcal{C}_{re}(\rho^{MS})=1.5$, while both reduce to unity at $\theta=0$ and $\theta=\frac{\pi}{2}$. This behavior indicates that the coherence detected in the chosen reference basis is strongly influenced by the relative weight of the components of the MS state. In contrast, the \textit{BICoher} remains constant, $\mathcal{C}_{BI}(\rho^{MS})\approx 0.84671$, over the entire range of $\theta$. Thus, although the conventional coherence measures vary substantially with $\theta$, the intrinsic coherence measured by \textit{BICoher} of the tripartite MS state remains unchanged. This distinction demonstrates that the $\theta$-dependent enhancement of $\mathcal{C}_{re}(\rho^{MS})$ and $\mathcal{C}_{\ell_1}(\rho^{MS})$ reflects the basis-sensitive manifestation of coherence rather than a modification of the underlying \textit{BICoher}. In particular, the invariance of $\mathcal{C}_{BI}(\rho^{MS})$ suggests that the family of MS states parametrized by $\theta$ possesses the same intrinsic coherence, despite exhibiting different amounts of coherence when quantified with respect to the selected basis.\\

\noindent The dependence of the coherence measures on $\theta$ can further be understood by comparing them with genuine tripartite entanglement. For the normalized maximally sliced state (MS) of Eq.(\ref{msstate}), the three-tangle is given by Eq.(\ref{tanglerhoMS}). Thus, $\tau(\rho^{MS})$ increases monotonically from zero at $\theta=0$ to its maximum value of unity at $\theta=\frac{\pi}{2}$. In contrast, both $\mathcal{C}_{re}$ and $\mathcal{C}_{\ell_1}$ are symmetric about $\theta=\frac{\pi}{4}$ and attain their maximum values at this balanced point. Therefore, the point of maximum basis-dependent coherence does not coincide with the point of maximum genuine tripartite entanglement. This difference provides a useful physical distinction between coherence and genuine tripartite entanglement in the MS state. At $\theta=0$, the three-tangle vanishes, $\tau(\rho^{MS})=0$, while the coherence measures remain finite. Hence, the presence of coherence in the state does not by itself imply the presence of genuine tripartite entanglement. As $\theta$ is varied in the range $\Big[0,~\frac{\pi}{2}\Big]$, genuine tripartite entanglement increases from $0$ to $1$, whereas the basis-dependent coherence first increases and subsequently decreases. This demonstrates that the two resources respond differently to the redistribution of amplitudes within the MS state. An additional feature is that the \textit{BICoher} remains fixed at $\mathcal{C}_{BI}\approx0.84671$ for the entire $\theta$ range. Consequently, the variations of $\mathcal{C}_{re}$ and $\mathcal{C}_{\ell_1}$ reflect changes in the way coherence is manifested with respect to the chosen basis, rather than a corresponding variations of the basis-independent coherent content. Taken together, these results show that the MS state provides a clear example in which basis-dependent coherence, \textit{BICoher}, and genuine tripartite entanglement exhibit distinct parameter dependence and therefore characterize different aspects of the underlying quantum correlations. If we express $\mathcal{C}_{\ell_{1}}$ and $\mathcal{C}_{re}$ for the state $\rho^{MS}$ in terms of tangle $\tau =\tau(\rho^{MS})$, then we get
\begin{eqnarray}
\label{cohertangle}
\mathcal{C}_{\ell_1}(\rho^{MS}) &=& \sqrt{1-\tau} + \sqrt{\tau} (1 + \sqrt{1-\tau})\\
\mathcal{C}_{re}(\rho^{MS}) &=& 1 - \tau\log_{2}\sqrt{\tau} - (1-\tau)\log_{2}\sqrt{1-\tau}.
\end{eqnarray}

\noindent We now discuss the characteristic features of the reduced bipartite MS state obtained by tracing out a party from it. 
\subsubsection{Bipartite MS state:}
\noindent Tracing out the third qubit from the tripartite MS state of Eq.~(\ref{msstate}) gives the reduced bipartite density matrix in Eq.~(\ref{densityMSAB}).
\begin{equation}
\label{densityMSAB}
\rho^{MS}_{AB} =
\left(
\begin{array}{cccc}
\frac{1}{2} & 0 & 0 & \frac{\cos\theta}{2} \\
0 & 0 & 0 & 0 \\
0 & 0 & 0 & 0 \\
\frac{\cos\theta}{2} & 0 & 0 & \frac{1}{2}
\end{array}
\right).
\end{equation}

\noindent Using Eqs. (\ref{l1norm}), (\ref{cr2}) and (\ref{BI}) we calculate the relative entropy, $\ell_1$-norm and \textit{BICoher} of coherence and obtain the following.

\begin{equation}
\label{allcoherences1}
\begin{aligned}
C_{\rm re}\left(\rho_{AB}^{\rm MS}\right)
={}&1+\frac{1+\cos\theta}{2}
\log_2\left(\frac{1+\cos\theta}{2}\right)\\
&+\frac{1-\cos\theta}{2}
\log_2\left(\frac{1-\cos\theta}{2}\right),\\[4pt]
C_{\ell_1}\left(\rho_{AB}^{\rm MS}\right)
={}&\cos\theta,\\[4pt]
C_{\rm BI}\left(\rho_{AB}^{\rm MS}\right)
={}&\Bigg[
-a\log_2a-b\log_2b
-\frac14\log_2\left(\frac18\right)\\
&\qquad
-1+\frac12(e)
\Bigg]^{\frac{1}{2}}.
\end{aligned}
\end{equation}

\noindent where 

\begin{eqnarray}
\label{abcd}
a &=& \frac{3-2\cos(\theta)}{8},~~ b = \frac{3+2\cos(\theta)}{8}\nonumber\\
c &=& \frac{1+\cos(\theta)}{2}, ~~ d = \frac{1-\cos(\theta)}{2}\nonumber\\
e &=& c\log_2 c + d\log_2 d.
\end{eqnarray}

\noindent We now plot all the coherences of Eq.(\ref{allcoherences1}) below.\\

\begin{figure}[htbp]
    \centering
    \includegraphics[width=0.98\columnwidth]{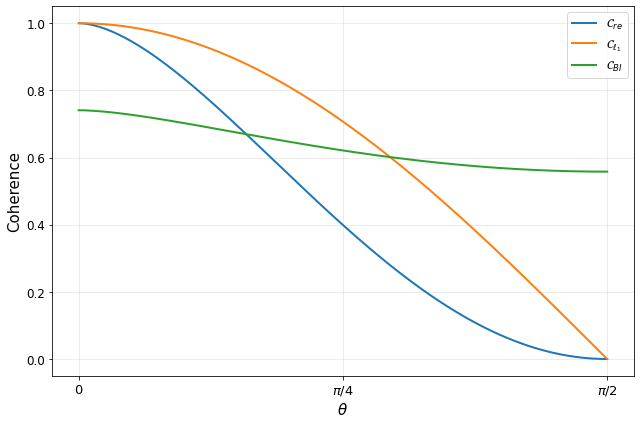}
    \caption{Coherence measures of the bipartite MS state as functions of $\theta$.}
    \label{fig2}
\end{figure}

\noindent The variation of the three coherence measures, namely $\mathcal{C}_{re}$, $\mathcal{C}_{\ell_{1}}$, and \textit{BICoher} ($\mathcal{C}_{BI}$), with the parameter $\theta$ is shown in Figure~\ref{fig2}. All three measures exhibit a decreasing behavior over the interval $0\leq\theta\leq \frac{\pi}{2}$, although their rates of decrease and their quantitative values are markedly different. This distinct behaviour highlights the different aspects of quantum coherence of the bipartite form of maximally sliced state $\rho^{MS}_{AB}$ captured by these three measures. At $\theta=0$, both basis-dependent coherence measures attain their maximum value, i.e. $\mathcal{C}_{re}(\rho^{MS}_{AB})=\mathcal{C}_{\ell_{1}}(\rho^{MS}_{AB})=1$. As $\theta$ increases, $\mathcal{C}_{re}$ decreases more rapidly than $\mathcal{C}_{\ell_{1}}$. In particular, at the balanced point $\theta=\frac{\pi}{4}$, the relative entropy of coherence $\mathcal{C}_{re}$ becomes $0.39910$ whereas the $\ell_{1}$-norm coherence remains finite, $\mathcal{C}_{\ell_{1}}$ is $0.70710$. Thus, the two basis-dependent measures do not provide the same quantitative characterization of coherence even though they are evaluated in the same computational basis.\\

\noindent   The $\ell_{1}$-norm coherence decreases from unity at $\theta=0$ to zero at $\theta=\frac{\pi}{2}$. In contrast, the basis-independent coherence $\mathcal{C}_{BI}$ for the MS state $\rho^{MS}_{AB}$ remains finite throughout the entire parameter range. It decreases from approximately $0.7408$ at $\theta=0$, passes through approximately $0.6211$ at $\theta=\frac{\pi}{4}$, and approaches approximately $0.5579$ at $\theta=\frac{\pi}{2}$. Therefore, even at $\theta=\frac{\pi}{2}$, where the computational-basis $\ell_{1}$ norm coherence vanishes, a finite amount of basis-independent coherence persists. The expression for $\mathcal{C}_{BI}$, involving the quantities $a$, $b$, $c$, $d$ and $e$ defined in Eq.~(\ref{allcoherences1}), explicitly demonstrates that this measure has a different dependence on $\theta$ from the conventional coherence measures. The disappearance of computational-basis coherence at the endpoint should not be interpreted as the complete disappearance of the coherent content of the bipartite MS state. Rather, the figure illustrates the important distinction between basis-dependent coherence and intrinsic basis-independent coherence.\\

\noindent If we now examine the entanglement structure of the bipartite state $\rho^{MS}_{AB}$, we find from Eq.~(\ref{co1}) that its concurrence has the same expression as its $\ell_1$-norm coherence, namely $\cos(\theta)$. This equality can be understood from the $X$-state structure of the density matrix. We know that a general two-qubit $X$ state is of the form \cite{nandi2018, rau2009, quesada2012, roy2024}
\begin{eqnarray}
\label{xstate}
\rho^{X} = \left(%
\begin{array}{cccc}
\rho_{11}	 &  0 & 0 & \rho_{14} \\
0 &  \rho_{22} & \rho_{23} & 0 \\
0	 &  \rho_{32} & \rho_{33} & 0\\
\rho_{41} & 0 & 0 & \rho_{44}\\
\end{array}%
\right),
\end{eqnarray}

\noindent whose concurrence is given by 
\begin{eqnarray}
\label{rhoxconc}
\mathcal{C}(\rho^{X}) = 2~\max[0,~|\rho_{14}| - \sqrt{\rho_{22}\rho_{33}},~|\rho_{23}| - \sqrt{\rho_{11}\rho_{44}}]\nonumber\\
\end{eqnarray}
\noindent Now for the reduced density matrix given in Eq.(\ref{densityMSAB}), the term $\rho_{14} = \frac{\cos(\theta)}{2}$, while $\rho_{22} = \rho_{33} = \rho_{23} = 0$, so that concurrence is $\mathcal{C}(\rho^{X}) = \cos(\theta)$. However, $\mathcal{C}_{\ell_1}(\rho^{X}) = 2~|\rho_{14}|$ and this is equal to $2~|\frac{\cos(\theta)}{2}| = \cos(\theta) = \mathcal{C}(\rho^{X})$ where $0\leq \theta \leq \frac{\pi}{2}$. Thus, the equality is essentially a consequence of the fact that the same single off-diagonal coherence is responsible for the entanglement of this particular two-qubit state.\\

\noindent For the reduced bipartite MS state, the correlation matrix is diagonal and its singular values are $\cos\theta$, $\cos\theta$, and $1$. Hence
\begin{equation}
\label{telepfidrhox}
N\left(\rho^{MS}_{AB}\right)=1+2\cos\theta,
\end{equation}
and the optimal teleportation fidelity is
\begin{equation}
\label{telepfidMS}
f^{T}_{\mathrm{opt}}\left(\rho^{MS}_{AB}\right)
=\frac{2+\cos\theta}{3}.
\end{equation}
Thus, the state is useful for standard teleportation whenever
$N(\rho^{MS}_{AB})>1$, which for the physical interval
$0\leq\theta\leq\frac{\pi}{2}$ gives $0\leq\theta<\frac{\pi}{2}$.\\

\noindent We would like to make a note here. For the MS state given in Eq.(\ref{msstate}), reducing the state to its bipartite form by tracing out the third party qubit will be useful as teleportation channel as has been described above. But if we remove the first qubit or second qubit from this maximally sliced state, it is found that the concurrence of the reduced bipartite states is 0 (zero) always (the reduced states are also not in the form of bipartite $X$ state) and hence the states are not useful for performing teleportation.\\

\noindent We shall now discuss the \textit{post-noise scenario}, where the bipartite state would be decohered by passing the state though the \textit{amplitude damping channel}. 

\section{Amplitude damping channel}
\label{sec:adc}

\noindent The amplitude damping (AD) channel is an important model for studying the dynamics of quantum resources in realistic quantum systems, as it describes the irreversible loss of excitation from a quantum system to its surrounding environment \cite{nc2000}. In physical settings, such a process is associated with energy relaxation and spontaneous emission, and therefore represents a fundamental decoherence mechanism in quantum information processing \cite{yueber2004}. The study of the MS state under amplitude damping is particularly relevant because the state contains superpositions of computational-basis states with different excitation numbers, making it sensitive to excitation loss. Consequently, the AD channel provides a natural framework for examining how quantum coherence, bipartite entanglement, and teleportation performance evolve when the system interacts with a dissipative environment. For a single qubit, the amplitude damping channel is described by the following Kraus operators:
\begin{eqnarray}
\label{kraussad}
E_0=
\begin{pmatrix}
1 & 0\\
0 & \sqrt{1-\gamma}
\end{pmatrix},
\qquad
E_1=
\begin{pmatrix}
0 & \sqrt{\gamma}\\
0 & 0
\end{pmatrix},
\end{eqnarray}
\noindent where $0\leq\gamma\leq1$ is the damping parameter. The parameter $\gamma$ characterizes the probability of excitation loss, with $\gamma=0$ corresponding to the absence of damping and $\gamma=1$ representing complete amplitude damping. The Kraus operators satisfy the completeness relation

\begin{eqnarray}
\label{krausscond1}
E_0^{\dagger}E_0+E_1^{\dagger}E_1=I,
\end{eqnarray}

\noindent which ensures that the resulting quantum operation is trace-preserving.\\

\noindent For the bipartite MS state of Eq.(\ref{densityMSAB}), we consider independent amplitude damping of the two constituent qubits. The corresponding two-qubit Kraus operators are therefore given by\
\begin{eqnarray}
\label{krausscond2}
K_{ij}=E_i\otimes E_j,
\qquad i,j\in\{0,1\}.
\end{eqnarray}

\noindent The action of the amplitude damping channel on the bipartite density matrix ($\rho$) is then described by the operator-sum representation

\begin{eqnarray}
\label{krausscond3}
\rho^{AD}
=
\sum_{i,j=0}^{1}
K_{ij}\rho K_{ij}^{\dagger}.
\end{eqnarray}

\noindent Applying the above channel to the bipartite MS state given in Eq.~(\ref{densityMSAB}), we obtain the post-noise density matrix
\begin{equation}
\label{postnoisestateadms}
\rho_{AB}^{\mathrm{MS}}\big|_{\mathrm{AD}}
=
\frac{1}{2}
\left(
\begin{smallmatrix}
1+\gamma^2 & 0 & 0 & (1-\gamma)\cos\theta \\
0 & \gamma(1-\gamma) & 0 & 0 \\
0 & 0 & \gamma(1-\gamma) & 0 \\
(1-\gamma)\cos\theta & 0 & 0 & (1-\gamma)^2
\end{smallmatrix}
\right).
\end{equation}

\noindent The density matrix (\ref{postnoisestateadms}) represents the decohered bipartite maximally sliced state after both qubits interact independently with dissipative environments through the amplitude damping channel. This post-noise state will subsequently be used to investigate the evolution of quantum coherence, entanglement, and teleportation performance under amplitude damping.\\

\subsection{Coherences after amplitude damping}
\noindent Using Eqs. (\ref{cr1}), (\ref{l1norm}) and (\ref{BI}), we obtain the relative entropy, $\ell_1$-norm, basis-independent coherences of the state $\rho_{AB}^{MS}|_{\mathrm{AD}}$.\\

\noindent Thus we have
\begin{eqnarray}
\label{allcoherencesad}
\mathcal{C}_{re}(\rho_{AB}^{MS}|_{AD})
&=& S(\rho_{d}^{MS}|_{AD}) - S(\rho_{AB}^{MS}|_{AD}),\nonumber\\
\mathcal{C}_{\ell_{1}}(\rho_{AB}^{MS}|_{AD})
&=& (1-\gamma)\cos\theta,\nonumber\\
\mathcal{C}_{BI}(\rho_{AB}^{MS}|_{AD})
&=& \sqrt{
S\Big(\frac{\rho_{AB}^{MS}|_{AD}+\frac{I}{4}}{2}\Big)
-\frac{S(\rho_{AB}^{MS}|_{AD})+2}{2}
}.\nonumber\\
\end{eqnarray}

\noindent where

\begin{eqnarray}
\label{crrhomsad1}
S(\rho_{d}^{MS}|_{AD}) &=& -\sum_{x}f_x~\log_{2}(f_x),\nonumber\\
f_1 &=& \frac{1}{2} - \gamma - \frac{\gamma^2}{2},\nonumber\\
f_2 &=& \frac{\gamma^2 + 1}{2},\nonumber\\
f_3 &=& \frac{\gamma - \gamma^2}{2} = f_4,
\end{eqnarray}

\begin{eqnarray}
\label{crrhomsad2}
S(\rho_{AB}^{MS}|_{AD}) &=& -\sum_{y}e_y~\log_{2}(e_y),\nonumber\\
e_1 &=& \frac{\gamma^2 - \gamma + 1}{2} + \frac{\sqrt{\cos^2(\theta)(\gamma - 1)^2 + \gamma^2}}{2},\nonumber\\
e_2 &=& \frac{\gamma^2 - \gamma + 1}{2} - \frac{\sqrt{\cos^2(\theta)(\gamma - 1)^2 + \gamma^2}}{2},\nonumber\\
e_3 &=& \frac{\gamma - \gamma^2}{2} = e_4.
\end{eqnarray}

\noindent Again, in the expression $\mathcal{C}_{BI}(\rho_{AB}^{MS}|_{AD})$ of Eq.~(\ref{allcoherencesad}), we have
\begin{equation}
\label{crrhomsadBI}
\resizebox{0.92\columnwidth}{!}{$
\begin{aligned}
S\Bigg(\frac{\rho_{AB}^{MS}|_{AD}+\frac{I}{4}}{2}\Bigg)
&= -\sum_{z}r_z\log_{2}(r_z),\\
r_1 &= \frac{2\gamma^2-2\gamma+3}{8}
+\frac{\sqrt{(1-\gamma)^2\cos^2\theta+\gamma^2}}{4},\\
r_2 &= \frac{2\gamma^2-2\gamma+3}{8}
-\frac{\sqrt{(1-\gamma)^2\cos^2\theta+\gamma^2}}{4},\\
r_3 &= \frac{1+2\gamma-2\gamma^2}{8}=r_4.
\end{aligned}
$}
\end{equation}

\subsubsection{Comparative analysis of coherence measures under amplitude damping}

\begin{figure}[htbp]
    \centering
    \includegraphics[width=0.95\columnwidth]{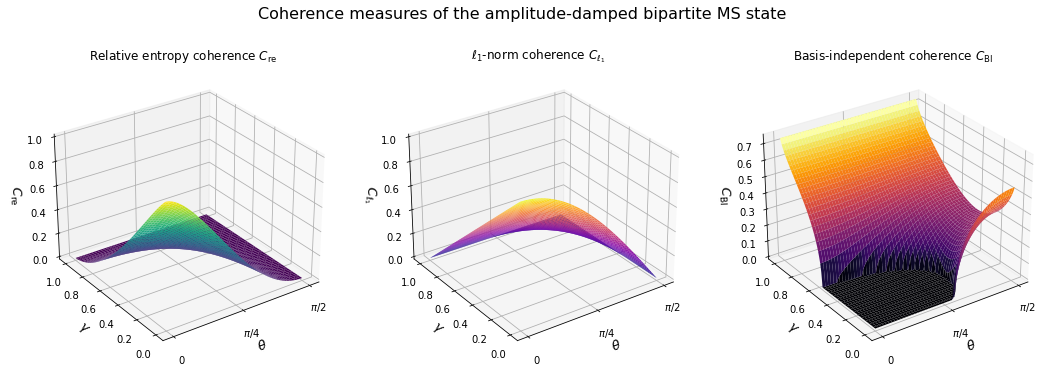}
    \caption{Three-dimensional plots of the relative entropy of coherence
    $\mathcal{C}_{\mathrm{re}}$, $\ell_{1}$-norm coherence $C_{\ell_1}$, and
    basis-independent coherence $\mathcal{C}_{\mathrm{BI}}$ of the amplitude-damped
    bipartite maximally sliced (MS) state
    $\rho^{\mathrm{MS}}_{AB|_{\mathrm{AD}}}$ as functions of the state
    parameter $\theta$ and the amplitude-damping parameter $\gamma$.}
    \label{fig3}
\end{figure}

\noindent Figure~\ref{fig3} presents the three-dimensional variation of the relative entropy of coherence $\mathcal{C}_{\mathrm{re}}$, the $\ell_1$-norm coherence $\mathcal{C}_{\ell_1}$, and the basis-independent coherence $\mathcal{C}_{\mathrm{BI}}$ as functions of the state parameter $\theta$ and the amplitude-damping parameter $\gamma$ of the state $\rho_{AB}^{MS}|_{\mathrm{AD}}$. The three surfaces exhibit distinctly different dependencies on these two parameters, demonstrating that the three coherence measures capture different aspects of the quantum coherence of the amplitude-damped bipartite maximally sliced (MS) state of Eq.(\ref{densityMSAB}). The $\ell_1$-norm coherence exhibits the simplest dependence on the two parameters. From Eq.~(\ref{allcoherencesad}), it is given by $(1-\gamma)\cos\theta$. Consequently, $\mathcal{C}_{\ell_1}$ decreases monotonically with change in the values of the state parameter $\theta$ and that of the noise parameter $\gamma$. For a fixed value of $\gamma$, the coherence decreases from its maximum value at $\theta=0$ to zero at $\theta=\frac{\pi}{2}$. Similarly, for a fixed $\theta$, the $\ell_1$-norm coherence decreases linearly with the damping parameter and vanishes at $\gamma=1$. This behaviour directly reflects the suppression of the off-diagonal element of the density matrix by the amplitude-damping channel. Thus, among the three measures considered, $\mathcal{C}_{\ell_1}$ provides the most direct representation of the loss of computational-basis coherence induced by the dissipative environment.\\

\noindent On the other hand, the relative entropy of coherence $\mathcal{C}_{\mathrm{re}}(\rho_{AB}^{MS}|_{AD})$ displays a more nonlinear dependence on both $\theta$ and $\gamma$. Unlike the $\ell_1$-norm, $\mathcal{C}_{\mathrm{re}}$ is determined by the difference between the von Neumann entropy of the diagonal part of the density matrix and that of the complete density matrix. Hence, its variation depends not only on the magnitude of the off-diagonal element but also on the redistribution of the eigenvalues of the state under amplitude damping. As seen in Figure~\ref{fig3}, $\mathcal{C}_{\mathrm{re}}$ decreases substantially as the damping parameter increases. Its dependence on $\theta$ is also non-linear, and the surface is distinctly different from that of $\mathcal{C}_{\ell_1}$. This difference emphasizes that the two basis-dependent measures, although evaluated in the same computational basis, do not quantify coherence in an equivalent manner.\\

\noindent A qualitatively different behaviour is exhibited by the basis-independent coherence $\mathcal{C}_{\mathrm{BI}}$ for the state $\rho_{AB}^{MS}|_{\mathrm{AD}}$. In contrast to the two basis-dependent measures, $\mathcal{C}_{\mathrm{BI}}$ is not defined with respect to a particular reference basis. Its dependence on $\theta$ and $\gamma$ therefore cannot be interpreted simply in terms of the magnitude of the computational-basis off-diagonal element. The surface in Figure~\ref{fig3} shows that $\mathcal{C}_{\mathrm{BI}}$ retains a finite value over the parameter domain, even in regions where the conventional computational-basis coherence becomes very small. In particular, the vanishing of $C_{\ell_1}$ at $\theta=\frac{\pi}{2}$ does not imply the disappearance of intrinsic coherence content of the state. This demonstrates the important distinction between the basis-sensitive manifestation of coherence and the basis-independent coherence (measured by \textit{BICoher}) of the quantum state.\\

\begin{figure}[htbp]
    \centering
    \includegraphics[width=0.4\textwidth]{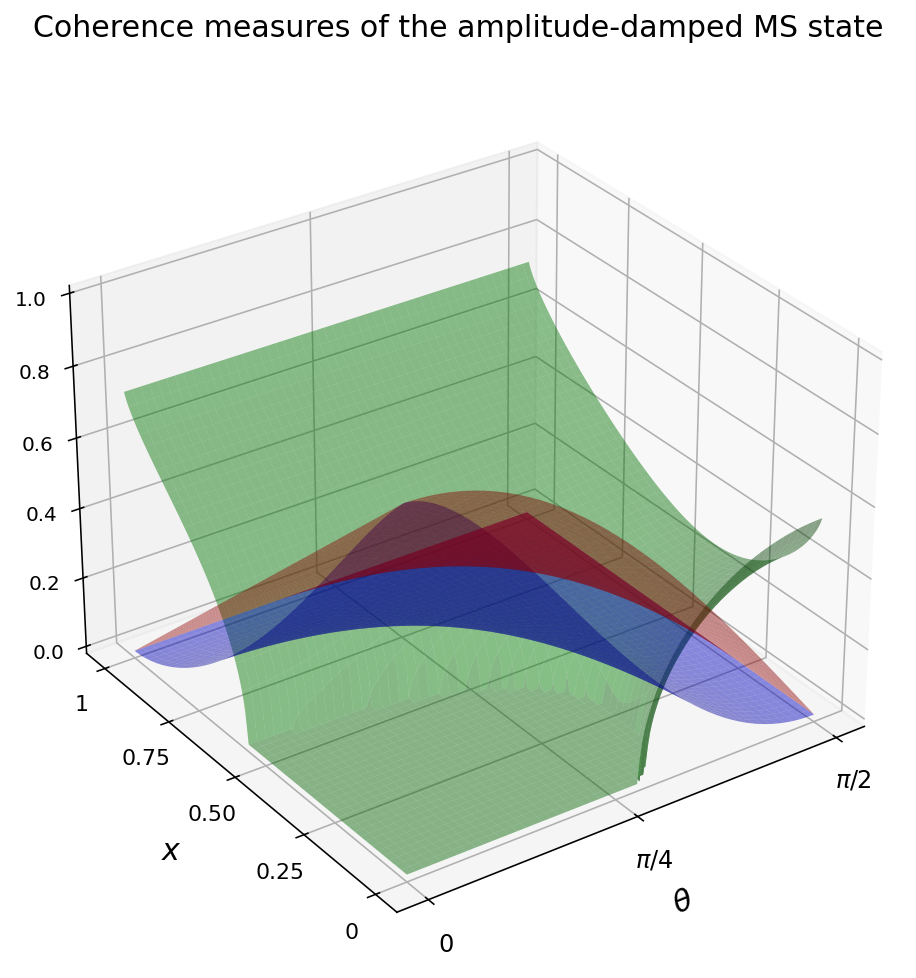}
    \caption{Combined three-dimensional representation of the relative entropy of coherence $\mathcal{C}_{\mathrm{re}}$, $\ell_{1}$-norm coherence $\mathcal{C}_{\ell_1}$, and basis-independent coherence $\mathcal{C}_{\mathrm{BI}}$ of the amplitude-damped bipartite maximally sliced (MS) state $\rho^{\mathrm{MS}}_{AB|\mathrm{AD}}$ as functions of $\theta$ and $\gamma$. }
    \label{fig4}
\end{figure}

\noindent A comparison of the three surfaces therefore reveals that amplitude damping affects the different coherence measures in qualitatively different ways. The $\ell_1$-norm coherence follows directly the suppression of the off-diagonal density-matrix element, whereas the relative entropy of coherence incorporates the changes in the complete spectral structure of the density matrix (which is clear from Figure~\ref{fig4}). The basis-independent coherence, on the other hand, provides a measure of coherent content that is not tied to the computational basis and consequently exhibits a different dependence on both $\theta$ and $\gamma$. The three-dimensional representation thus demonstrates that the dissipative evolution of the MS state cannot be characterized by a single coherence measure. Rather, $\mathcal{C}_{\mathrm{re}}$, $\mathcal{C}_{\ell_1}$, and $\mathcal{C}_{\mathrm{BI}}$ provide complementary descriptions of how quantum coherence evolves with the state parameter and the strength of amplitude damping.\\

\noindent \subsection{Concurrence and teleportation fidelity under amplitude damping}
\noindent For the post-noise state given in Eq.~(\ref{postnoisestateadms}), the concurrence can be obtained using Eq.~(\ref{rhoxconc}). Since $\rho_{AB|\mathrm{AD}}^{\mathrm{MS}}$ is an $X$ state, its concurrence takes the form

\begin{equation}
\label{concrhoadms}
\begin{aligned}
C\left(\rho_{AB|\mathrm{AD}}^{\mathrm{MS}}\right)
&=
2\max\left\{
0,
\frac{(1-\gamma)\cos\theta}{2}
-\frac{\gamma(1-\gamma)}{2}
\right\}\\
&=(1-\gamma)\max\{0,\cos\theta-\gamma\}.
\end{aligned}
\end{equation}

\noindent Thus, the amplitude-damped MS state possesses nonzero bipartite entanglement only when
\begin{equation}
\cos\theta>\gamma.
\end{equation}

\noindent Equivalently, the entangled region is
\begin{equation}
0\leq\theta<\cos^{-1}(\gamma),
\qquad
0\leq\gamma<1.
\end{equation}
\noindent The boundary $\gamma=\cos\theta$ therefore marks the sudden death of concurrence.\\

\noindent We next consider the teleportation fidelity of the amplitude-damped state. For a general two-qubit state, the optimal teleportation fidelity is given by Eq.~(\ref{telepfid1}). For the present $X$ state, the correlation matrix is diagonal, with singular values
\[
(1-\gamma)\cos\theta,\qquad
(1-\gamma)\cos\theta,\qquad
1-2\gamma+2\gamma^2.
\]
Consequently,

\begin{equation}
\label{telepAD}
\begin{aligned}
N\left(\rho_{AB|\mathrm{AD}}^{\mathrm{MS}}\right)
&=
2(1-\gamma)\cos\theta
+1-2\gamma+2\gamma^2,
\\[2mm]
f_{\mathrm{opt}}^T
\left(\rho_{AB|\mathrm{AD}}^{\mathrm{MS}}\right)
&=
\frac{1}{2}
\left[
1+
\frac{
1-2\gamma+2\gamma^2
+2(1-\gamma)\cos\theta
}{3}
\right]
\\
&=
\frac{
4-2\gamma+2\gamma^2
+2(1-\gamma)\cos\theta
}{6}.
\end{aligned}
\end{equation}

The classical teleportation fidelity is $2/3$. Therefore,
\begin{equation}
\begin{aligned}
f_{\mathrm{opt}}^T>\frac{2}{3}
&\Longleftrightarrow
2(1-\gamma)(\cos\theta-\gamma)>0\\
&\Longleftrightarrow
\cos\theta>\gamma,
\qquad 0\leq\gamma<1.
\end{aligned}
\end{equation}

\noindent Hence, for the amplitude-damped MS state, the condition for a nonzero concurrence and that for exceeding the classical
teleportation threshold coincide. The physically relevant region in which the state retains bipartite entanglement and simultaneously serves as a quantum teleportation channel is therefore

\begin{equation}
0\leq\theta<\frac{\pi}{2},
\qquad
0\leq\gamma<\cos\theta.
\end{equation}

\noindent It is also useful to compare the teleportation fidelity with its value in the absence of noise. As discussed above, in the pre-noise scenario, the teleportation fidelity has been obtained and let this teleportation fidelity be denoted by

\begin{equation}
f_0^T=\frac{2+\cos\theta}{3},
\end{equation}
whereas the amplitude-damped value is given by Eq.~(37). Their
difference is
\begin{equation}
f_0^T-f_{\mathrm{opt}}^T
=
\frac{\gamma(1-\gamma+\cos\theta)}{3}\geq0,
\qquad
0\leq\gamma\leq1.
\end{equation}
\noindent Thus, amplitude damping progressively reduces the teleportation fidelity. In the limit $\gamma\rightarrow1$, the state approaches a classical product state and $f_{\mathrm{opt}}^T\rightarrow \frac{2}{3}$.\\

\section{Dephasing channel}
\label{sec:dp}
\noindent  Having studied the effect of amplitude damping on the bipartite maximally sliced (MS) state, we now investigate the effect of dephasing noise on the same state. The study of dephasing is important because it represents a qualitatively different decoherence mechanism from amplitude damping. While amplitude damping describes the irreversible loss of excitation through energy relaxation, dephasing primarily affects the relative phases between the components of a quantum superposition without directly changing their populations. Since quantum coherence and the nonlocal correlations required for quantum information processing are sensitive to phase fluctuations, dephasing provides an important model for studying the degradation of quantum resources in realistic environments. In particular, comparing the behaviour of the bipartite MS state under amplitude damping and
dephasing allows us to distinguish the effects arising from population relaxation from those caused purely by the loss of phase coherence \cite{nc2000, breuer2002theory}.\\

\noindent For a single qubit, the dephasing channel can be represented by the operator-sum representation
\begin{eqnarray}
\label{dpsinglequbit}
    \rho_{\mathrm{DP}}
    =
    \sum_{i=0}^{1} F_i \rho F_i^\dagger,
\end{eqnarray}

\noindent where the Kraus operators are given by

\begin{eqnarray}
\label{dpkraus}
    F_0 =
    \sqrt{1-\frac{\kappa}{2}}\,I,
    \qquad
    F_1 =
    \sqrt{\frac{\kappa}{2}}\,\sigma_z,
    \qquad
    0 \leq \kappa \leq 1.
\end{eqnarray}

\noindent Here, \(I\) denotes the two-dimensional identity matrix and
\(\sigma_z\) is the Pauli-\(z=\) matrix, i.e. $\sigma_z =
    \begin{pmatrix}
        1 & 0\\
        0 & -1
    \end{pmatrix}$.

\noindent The parameter \(\kappa\) characterizes the strength of the dephasing process. The limiting case \(\kappa=0\) corresponds to the absence of dephasing, whereas increasing \(\kappa\) leads to a progressive loss of phase coherence. For the present parametrization, \(\kappa=1\) corresponds to complete dephasing of the off-diagonal elements of a single-qubit density matrix. The Kraus operators satisfy the completeness relation $F_0^\dagger F_0 + F_1^\dagger F_1 = I$ which ensures that the dephasing map is trace preserving.  For the bipartite MS state, we consider independent dephasing of the two constituent qubits. 
The action of the independent dephasing channel on the bipartite density matrix is described by the operator-sum representation
\begin{eqnarray}
\label{krausstwodp}
    \rho_{\mathrm{DP}}
    =
    \sum_{i,j=0}^{1}
    K_{ij}\rho K_{ij}^{\dagger}.
\end{eqnarray}

\noindent Since the dephasing channel acts independently on both qubits, the two-qubit coherence between the states \(\ket{00}\) and \(\ket{11}\) is suppressed by the factor \((1-\kappa)^2\). Importantly, the populations of the computational-basis states remain unchanged. Therefore, for the bipartite MS state of Eq.~(\ref{densityMSAB}), the dephasing process modifies the off-diagonal terms while preserving the diagonal terms. Applying the above channel to the bipartite MS state of Eq.~(\ref{densityMSAB}), we obtain the post-dephasing density matrix as

\begin{equation}
\label{densitydpms}
\rho^{\mathrm{MS}}_{AB|\mathrm{DP}}
=
\frac{1}{2}
\begin{pmatrix}
1 & 0 & 0 & (1-\kappa)^2\cos\theta\\
0 & 0 & 0 & 0\\
0 & 0 & 0 & 0\\
(1-\kappa)^2\cos\theta & 0 & 0 & 1
\end{pmatrix}.
\end{equation}

\subsection{Coherence measures after dephasing}

\noindent Using Eqs.~(\ref{cr1}), (\ref{l1norm}), and (\ref{BI}), we calculate the relative entropy of coherence, $\ell_1$-norm coherence, and basis-independent coherence of the dephased bipartite MS state $\rho^{\mathrm{MS}}_{AB|\mathrm{DP}}$.

\begin{equation}
\label{eq:dephasing_coherences}
\resizebox{0.82\columnwidth}{!}{$
\begin{aligned}
\mathcal{C}_{\rm re}\!\left(\rho^{\mathrm{MS}}_{AB|\mathrm{DP}}\right)
={}&1
+\frac{1+q}{2}\log_{2}\!\left(\frac{1+q}{2}\right)
+\frac{1-q}{2}\log_{2}\!\left(\frac{1-q}{2}\right),\\[4pt]
\mathcal{C}_{\ell_1}\!\left(\rho^{\mathrm{MS}}_{AB|\mathrm{DP}}\right)
={}&q,\\[4pt]
\mathcal{C}_{\rm BI}\!\left(\rho^{\mathrm{MS}}_{AB|\mathrm{DP}}\right)
={}&\Bigg[
-r_-\log_2 r_-\\
&\quad-r_+\log_2 r_+
-\frac14\log_2\!\left(\frac18\right)\\
&\quad-\frac{S_{\mathrm{DP}}+2}{2}
\Bigg]^{1/2}.
\end{aligned}
$}
\end{equation}

\noindent where
\begin{equation}
q=(1-\kappa)^2\cos\theta,
\qquad
r_{\pm}=\frac{3\pm2q}{8}.
\end{equation}
Here $S_{\mathrm{DP}}\equiv S(\rho^{\mathrm{MS}}_{AB|\mathrm{DP}})$, and
\begin{equation}
\label{Sdephased}
\begin{aligned}
S\left(\rho^{\mathrm{MS}}_{AB|\mathrm{DP}}\right)
={}&-\frac{1+q}{2}\log_2\left(\frac{1+q}{2}\right)\\
&-\frac{1-q}{2}\log_2\left(\frac{1-q}{2}\right).
\end{aligned}
\end{equation}

\subsubsection{Comparative analysis of coherence measures under dephasing}

\begin{figure}[htbp]
    \centering
    \includegraphics[width=0.95\columnwidth]{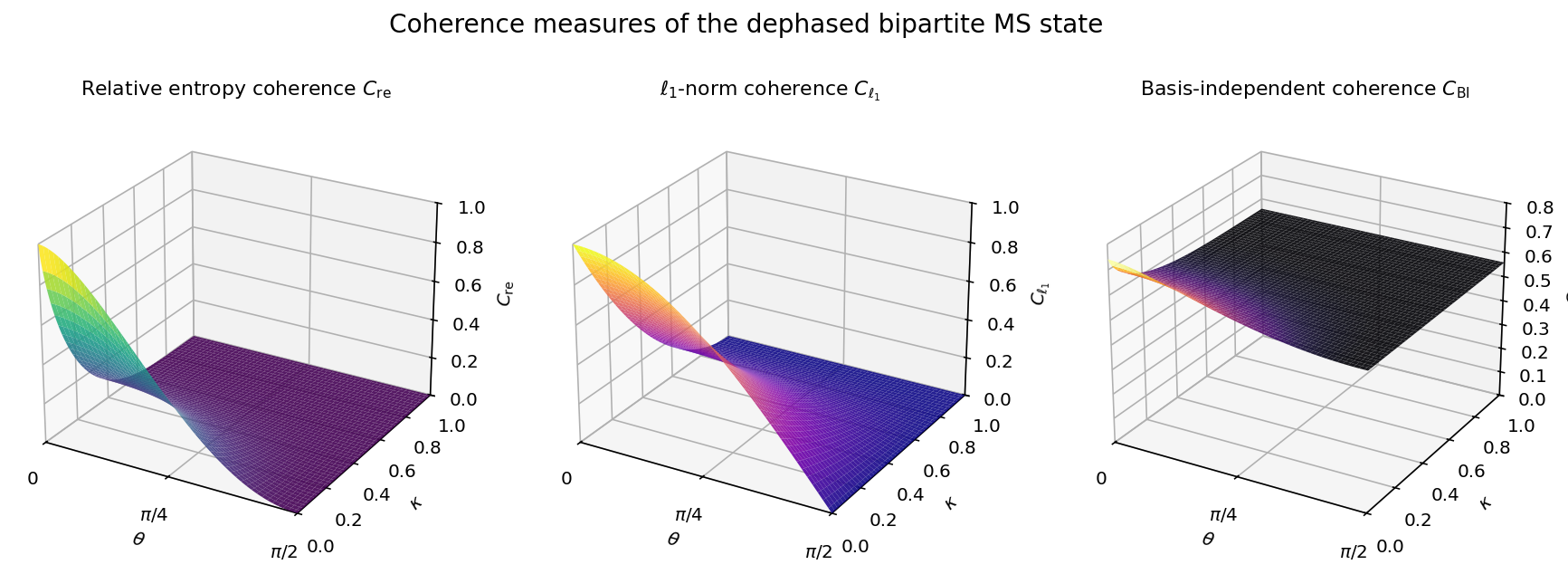}
    \caption{Three-dimensional plots of the relative entropy of coherence
    $\mathcal{C}_{\rm re}$, $\ell_1$-norm coherence $\mathcal{C}_{\ell_1}$, and
    basis-independent coherence $C_{\rm BI}$ of the dephased bipartite
    maximally sliced (MS) state
    $\rho^{\mathrm{MS}}_{AB|\mathrm{DP}}$ as functions of the state
    parameter $\theta$ and the dephasing parameter $\kappa$.}
    \label{fig5}
\end{figure}

\noindent Figure \ref{fig5} presents the three-dimensional variation of the relative entropy of coherence $\mathcal{C}_{\rm re}$, the $\ell_1$-norm coherence $\mathcal{C}_{\ell_1}$, and the basis-independent coherence $C_{\rm BI}$ as functions of the state parameter $\theta$ and the dephasing parameter $\kappa$ for the dephased bipartite maximally sliced (MS) state. The three surfaces exhibit distinct dependencies on the two parameters, demonstrating that the different coherence measures capture different aspects of
the decoherence induced by the dephasing channel.\\

\noindent The $\ell_1$-norm coherence exhibits the simplest dependence on the two parameters viz. $\kappa$ (the noise parameter) and $\theta$ (the state parameter). From Eq.~(\ref{eq:dephasing_coherences}), it is given by $C_{\ell_1}=(1-\kappa)^2\cos\theta$. Consequently, for a fixed value of $\kappa$, the $\ell_1$-norm coherence decreases monotonically with $\theta$, reaching zero at $\theta=\frac{\pi}{2}$. Similarly, for a fixed $\theta$, it decreases monotonically with increasing $\kappa$ and vanishes at $\kappa=1$. This behaviour directly reflects the suppression of the off-diagonal element of the density matrix by the dephasing channel. Since the populations remain unchanged under dephasing, the variation of $C_{\ell_1}$ provides a direct representation of the loss of computational-basis coherence caused by phase fluctuations.\\

\noindent The relative entropy of coherence $C_{\rm re}$ also decreases with increasing $\theta$ and $\kappa$. However, unlike the $\ell_1$-norm coherence, its dependence on the effective coherence parameter $q=(1-\kappa)^2\cos\theta$ is nonlinear, as follows directly from Eq.~(\ref{eq:dephasing_coherences}). Consequently, the $\mathcal{C}_{\rm re}$ surface exhibits a curved profile in the three-dimensional representation. This behaviour originates from the logarithmic dependence of $\mathcal{C}_{\rm re}$ on the eigenvalues of the density matrix, rather than from a direct proportionality to the magnitude of the off-diagonal element. In the present dephasing process, although the populations remain unchanged, the suppression of the off-diagonal terms modifies the eigenvalues of the density matrix through the parameter $q=(1-\kappa)^2\cos\theta$. Hence, the relative entropy of coherence does not simply follow the magnitude of the off-diagonal element, but also reflects the corresponding change in the spectral structure of the state.\\

\noindent The basis-independent coherence $\mathcal{C}_{\rm BI}$ exhibits a qualitatively different behaviour from the two basis-dependent measures. Although it decreases as the effective coherence parameter $q$ is suppressed, it remains finite even when the computational-basis coherence vanishes. In particular, at $\theta=\frac{\pi}{2}$ or $\kappa=1$, one has $q=0$ and consequently $\mathcal{C}_{\ell_1}=\mathcal{C}_{\rm re}=0$, whereas $\mathcal{C}_{\rm BI}$ remains finite. Thus, the vanishing of the conventional computational-basis coherence under complete dephasing does not imply the complete disappearance of the coherent content measured by the basis-independent quantity.\\

\begin{figure}[htbp]
    \centering
    \includegraphics[width=0.4\textwidth]{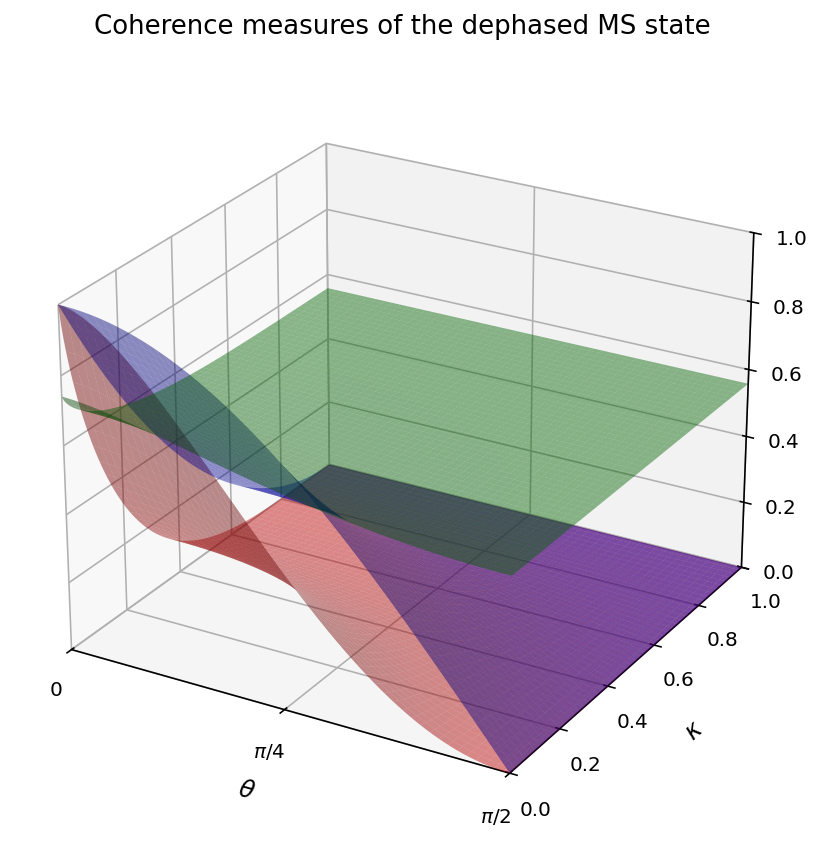}
    \caption{Combined three-dimensional representation of the relative
entropy of coherence $\mathcal{C}_{\rm re}$, $\ell_1$-norm coherence
$\mathcal{C}_{\ell_1}$, and basis-independent coherence $\mathcal{C}_{\rm BI}$ of the
dephased bipartite maximally sliced (MS) state
$\rho^{\mathrm{MS}}_{AB|\mathrm{DP}}$ as functions of $\theta$ and
$\kappa$. }
    \label{fig6}
\end{figure}

\noindent Figure~\ref{fig6} provides a combined three-dimensional representation of the relative entropy of coherence $\mathcal{C}_{\rm re}$, the $\ell_1$-norm coherence $\mathcal{C}_{\ell_1}$, and the basis-independent coherence $\mathcal{C}_{\rm BI}$ as functions of $\theta$ and $\kappa$. The combined surfaces allow a direct comparison of the responses of the three coherence measures to the increasing strength of dephasing. A common feature of the three surfaces is their dependence on the effective parameter $q=(1-\kappa)^2\cos\theta$. As either $\theta$ or $\kappa$ increases, $q$ decreases, leading to a progressive reduction of the basis-dependent coherence. The $\ell_1$-norm coherence follows this dependence directly through $\mathcal{C}_{\ell_1}=q$, and therefore forms the most straightforward surface in the combined representation. In contrast, $\mathcal{C}_{\rm re}$ exhibits a nonlinear variation because it depends on the entropy difference between the diagonal and complete density matrices. Thus, even though both measures decrease with the effective suppression of the off-diagonal coherence, they do not possess the same quantitative dependence.\\

\noindent The combined representation also clearly distinguishes the behaviour of the basis-independent coherence from that of the two basis-dependent measures. When $\theta=\frac{\pi}{2}$ or $\kappa=1$, the effective parameter $q$ becomes zero and both $\mathcal{C}_{\rm re}$ and $\mathcal{C}_{\ell_1}$ vanish. However, $\mathcal{C}_{\rm BI}$ remains finite. This feature emphasizes that the disappearance of coherence in the computational basis should not be interpreted as the complete absence of coherence as captured by the basis-independent measure \textit{BICoher}.

\subsection{Concurrence and teleportation fidelity under dephasing}

\noindent We now investigate the bipartite entanglement and teleportation performance of the dephased MS state of Eq.(\ref{densitydpms}). This state is an $X$ state, and hence its concurrence can be obtained using Eq.~(\ref{rhoxconc}). Thus
the concurrence is given as
\begin{align}
\label{concdephasedms}
\mathcal{C}\left(\rho^{\mathrm{MS}}_{AB|\mathrm{DP}}\right)
&=
2\max\left\{
0,\frac{q}{2}
\right\} \nonumber\\
&=(1-\kappa)^2\cos\theta,
\end{align}
\noindent where $q=(1-\kappa)^2\cos\theta$. 
Thus,
\begin{eqnarray}
\label{conccohdephasedms}
\mathcal{C}\left(\rho^{\mathrm{MS}}_{AB|\mathrm{DP}}\right)
=(1-\kappa)^2\cos\theta
=\mathcal{C}_{\ell_1}
\left(\rho^{\mathrm{MS}}_{AB|\mathrm{DP}}\right).
\end{eqnarray}

\noindent The equality between concurrence and $\ell_1$ norm coherence follows from the fact that the dephased MS state contains only one relevant off-diagonal pair, namely $\rho_{14}=\rho_{41}=\frac{q}{2}$, which is also the term responsible for the concurrence. We next evaluate the teleportation fidelity of the dephased MS state. Using Eqs.(\ref{telepfid1}) and (\ref{telepfid2}) we get  

\begin{eqnarray}
\label{dephsedmsn}
N\left(\rho^{\mathrm{MS}}_{AB|\mathrm{DP}}\right)
=
1+2q
=
1+2(1-\kappa)^2\cos\theta.
\end{eqnarray}

\noindent and 

\begin{align}
\label{dephasedtelep}
f_{\mathrm{opt}}^T
\left(\rho^{\mathrm{MS}}_{AB|\mathrm{DP}}\right)
&=
\frac{1}{2}
\left[
1+
\frac{1+2(1-\kappa)^2\cos\theta}{3}
\right]
\nonumber\\
&=
\boxed{
\frac{2+(1-\kappa)^2\cos\theta}{3}
}.
\end{align}

\noindent The condition for useful standard teleportation is $N(\rho)>1$. For the dephased MS state this gives $(1-\kappa)^2\cos\theta>0$. Therefore, within the physical parameter range $0\leq\theta\leq \frac{\pi}{2}$ and $0\leq\kappa\leq1$, the state is useful for standard teleportation for

\begin{eqnarray}
\label{conddephasedms}
0\leq\theta<\frac{\pi}{2},
\qquad
0\leq\kappa<1.
\end{eqnarray}

\noindent At $\theta=\frac{\pi}{2}$ or $\kappa=1$, one has $q=(1-\kappa)^2\cos(\theta)=0$, so that the concurrence vanishes and the teleportation fidelity reaches the classical value $\frac{2}{3}$. An important feature of the dephased MS state is that the teleportation fidelity can be expressed directly in terms of its concurrence:

\begin{eqnarray}
\label{telepconcdephasedrel1}
\boxed{
f_{\mathrm{opt}}^T=\frac{2+\mathcal{C}}{3}.
}
\end{eqnarray}

\noindent Thus, for this state, nonzero concurrence and teleportation fidelity above the classical threshold occur in the same parameter region. It is instructive to compare this result with the amplitude-damping case. For the amplitude-damped MS state, Eq.~(\ref{telepAD}) gives, in the region where the concurrence is nonzero,
\begin{equation}
\label{telepconcAD}
f_{\mathrm{opt}}^T
=
\frac{2+\mathcal{C}_{\mathrm{AD}}}{3},
\qquad
\mathcal{C}_{\mathrm{AD}}
=(1-\gamma)(\cos\theta-\gamma).
\end{equation}
Thus, for the present MS-derived channel, the optimal teleportation fidelity has the same concurrence dependence in the entangled regime for both noise models. The physical distinction between amplitude damping and dephasing is instead encoded in their different dependence of concurrence on the corresponding noise parameter.
\section{Comparison of teleportation performance under amplitude damping and dephasing}

\noindent The teleportation performance of the bipartite MS state is affected by amplitude damping and dephasing through different physical mechanisms. Amplitude damping describes energy relaxation and modifies both the populations and the off-diagonal coherence, whereas dephasing preserves the populations and suppresses phase coherence. For the present MS-derived channel, both noise models yield
$f_{\mathrm{opt}}^T=\frac{2+\mathcal{C}}{3}$ in the regime of nonzero concurrence. The difference therefore lies in how the two channels modify the concurrence: amplitude damping gives
$\mathcal{C}_{\mathrm{AD}}=(1-\gamma)(\cos\theta-\gamma)$, whereas dephasing gives
$\mathcal{C}_{\mathrm{DP}}=(1-\kappa)^2\cos\theta$. This comparison separates the common concurrence--fidelity relation from the distinct physical effects of dissipative and phase-decoherence processes.
\section{Conclusion}
\label{sec:conc}
\noindent We have investigated the quantum coherence, bipartite entanglement, and teleportation performance of the maximally sliced (MS) state, with particular emphasis on its behaviour under amplitude damping and dephasing. For the tripartite MS state, the basis-dependent coherence measures, namely the relative entropy of coherence and the $\ell_1$-norm coherence, exhibit a non-monotonic dependence on the state parameter $\theta$, whereas the basis-independent coherence remains invariant. In contrast, the three-tangle increases monotonically with $\theta$. These results demonstrate that basis-dependent coherence and genuine tripartite entanglement characterize different aspects of the quantum resources of the MS family.\\

\noindent Upon tracing out one of the parties, the resulting bipartite MS state retains entanglement and useful teleportation capability. An important feature of this reduced state is the equality between its concurrence and $\ell_1$-norm coherence, which arises because the same off-diagonal density-matrix elements determine both quantities. We have further studied the evolution of this bipartite state under amplitude damping and dephasing. The two noise channels affect the state in qualitatively different ways: amplitude damping modifies both populations and off-diagonal coherence, whereas dephasing preserves the populations and suppresses phase coherence. Consequently, the three coherence measures respond differently to the two noise processes. For the present MS-derived teleportation channel, the optimal teleportation fidelity satisfies $f_{\mathrm{opt}}^T=\frac{2+\mathcal{C}}{3}$ in the regime of nonzero concurrence for both channels. The distinct dependence of concurrence on $\gamma$ and $\kappa$ nevertheless reflects the different physical mechanisms of dissipative and phase decoherence. Overall, the results show that coherence, entanglement, and teleportation fidelity provide complementary descriptions of the quantum resources of the MS state and illustrate how the MS family can serve as a tunable framework for studying quantum-information resources under decoherence.

\vskip 0.5cm
\section*{Acknowledgements}
Sovik Roy acknowledges Dr.~Chandrashekar Radhakrishnan for introducing the measure of basis-independent coherence used in this work.

\section*{Author Contributions}
All authors contributed equally to the conception, formulation, analysis, writing, and interpretation of the present work.

\section*{Data Availability}
All data supporting the findings of this study are available within the paper.

\section*{Competing Interests}
The authors declare no competing interests.

\section*{Corresponding Author}
Correspondence should be addressed to Sovik Roy (s.roy2.tmsl@ticollege.org).

\end{document}